\documentclass[12pt]{article}
\usepackage{axodraw}
\usepackage{epsfig}
\def\e{{\rm e}}
\newcommand{\be}{\begin{equation}}
\newcommand{\ee}{\end{equation}}
\newcommand{\bea}{\begin{eqnarray}}
\newcommand{\eea}{\end{eqnarray}}

\newcommand{\gm}{\gamma}
\newcommand{\Gm}{\Gamma}

\newcommand{\ep}{\epsilon}

\newcommand{\dd}{\mbox{d}}

\newcommand{\nn}{\nonumber}

\begin{document}
\parindent=1.5pc

\begin{titlepage}
\rightline{hep-ph/0305142}
\rightline{May 2003}
\bigskip
\begin{center}
{{\bf Analytical Result for Dimensionally Regularized Massless
On-Shell Planar Triple Box
}} \\
\vglue 5pt
\vglue 1.0cm
{ {\large V.A. Smirnov\footnote{E-mail: smirnov@theory.sinp.msu.ru}
} }\\
\baselineskip=14pt
\vspace{2mm}
{\em Nuclear Physics Institute of Moscow State University}\\
{\em Moscow 119992, Russia}
\vglue 0.8cm
{Abstract}
\end{center}
\vglue 0.3cm
{\rightskip=3pc
 \leftskip=3pc
\noindent
The dimensionally regularized massless on-shell planar triple box
{}Feynman diagram with powers of propagators equal to one is
analytically evaluated for general values of the Mandelstam
variables $s$ and $t$ in a Laurent expansion in the parameter
$\ep=(4-d)/2$ of dimensional regularization up to a finite part.
An explicit result is expressed in terms of harmonic
polylogarithms, with parameters $0$ and $1$, up to the sixth order.
The evaluation is based on the method of Feynman parameters and
multiple Mellin--Barnes representation. The same technique can be
quite similarly applied to planar triple boxes with any numerators
and integer powers of the propagators.
\vglue 0.8cm}
\end{titlepage}

In the last four years, the problem of analytical evaluation was completely
solved for
most important classes of two-loop Feynman diagrams with
four external lines within dimensional regularization
\cite{dimreg}.
In the pure massless case with all end-points on-shell, i.e.
$p_i^2=0,\;i=1,2,3,4$, this was done in \cite{K1,SV,Tausk,AGO,ATT,GR3}.
The corresponding analytical algorithms were successfully
applied to the evaluation of various two-loop virtual
corrections  \cite{appl}.
In the case of massless two-loop four-point diagrams with
one leg off-shell the problem of evaluation was solved
in \cite{S2,GR2}, with subsequent applications \cite{appl3j} to the
process $e^+e^-\to 3$jets.
A first result for the massive on-shell case was presented in
\cite{S3}.
(See \cite{S4,S6} for  brief reviews of results on the
analytical evaluation of various double-box Feynman integrals
and the corresponding methods of evaluation.)

In \cite{S5,S6}, first analytical results on three-loop
on-shell massless four-point diagrams within dimensional regularization were
obtained.
The leading power asymptotic behaviour
of the dimensionally regularized massless
on-shell planar triple box diagram shown in Fig.~1
\begin {figure} [htbp]
\begin{picture}(400,80)(-140,-10)
\Line(-15,0)(0,0)
\Line(-15,50)(0,50)
\Line(165,0)(150,0)
\Line(150,0)(100,0)
\Line(150,50)(100,50)
\Line(165,50)(150,50)
\Line(0,0)(50,0)
\Line(50,0)(100,0)
\Line(100,50)(50,50)
\Line(50,50)(0,50)
\Line(0,50)(0,0)
\Line(50,0)(50,50)
\Line(100,0)(100,50)
\Line(150,0)(150,50)
\Vertex(0,0){1.5}
\Vertex(50,0){1.5}
\Vertex(100,0){1.5}
\Vertex(0,50){1.5}
\Vertex(50,50){1.5}
\Vertex(100,50){1.5}
\Vertex(150,0){1.5}
\Vertex(150,50){1.5}
\Text(-22,0)[]{$p_1$}
\Text(174,0)[]{$p_3$}
\Text(-22,50)[]{$p_2$}
\Text(174,50)[]{$p_4$}
\Text(25,43)[]{\small 1}
\Text(-7,25)[]{\small 2}
\Text(25,7)[]{\small 3}
\Text(75,7)[]{\small 4}
\Text(43,25)[]{\small 7}
\Text(93,25)[]{\small 5}
\Text(75,43)[]{\small 6}
\Text(125,7)[]{\small 8}
\Text(125,43)[]{\small 10}
\Text(143,25)[]{\small 9}
\end{picture}
\caption{Planar triple box diagram.}
\end{figure}
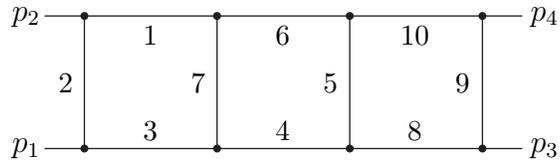
in the Regge limit $t/s \to 0$ was analytically evaluated in \cite{S5}
with the help of the strategy of expansion by regions \cite{BS}.
Then, in \cite{S6}, explicit analytical results for the unexpanded
master planar triple box were presented for  $1/\ep^j$ terms of
Laurent expansion in $\ep$ with j=6,5,4,3 and~2.

The purpose of this paper is to complete this task i.e. analytically evaluate
the missing $1/\ep$ part and the finite part.
An explicit result will be expressed in terms of harmonic polylogarithms (HPL)
\cite{HPL}, with parameters $0$ and $1$, up to the sixth order.
The evaluation is based on the technique of
alpha parameters and Mellin--Barnes (MB) representation
which was successfully used in \cite{K1,Tausk,S2,S3} and reduces,
due to taking residues and shifting contours, to
a decomposition of a given MB integral into pieces
where a Laurent expansion of the integrand in $\ep$ becomes possible.

The general planar triple box Feynman integral without numerator
takes the form
\bea
T(a_1,\ldots,a_{10};s,t;\ep) &=&
\int\int\int \frac{\dd^dk \, \dd^dl \, \dd^dr}{(k^2)^{a_1}
[(k+p_2)^2]^{a_2}
[(k+p_1+p_2)^2]^{a_3}}
\nn \\ && \hspace*{-10mm}
\times \frac{1}{
[(l+p_1+p_2)^2]^{a_4}[(r-l)^2]^{a_5}
(l^2)^{a_6} [(k-l)^2]^{a_7} }
\nn \\ && \hspace*{-10mm}
\times \frac{1}{
[(r+p_1+p_2)^2]^{a_8} [(r+p_1+p_2+p_3)^2]^{a_9}
(r^2)^{a_{10}} }
\, ,
\label{3box}
\eea
where $s=(p_1+p_2)^2$ and $t=(p_2+p_3)^2$ are Mandelstam variables and
$k,l,r$ are the loop momenta.
Usual prescriptions $k^2=k^2+i 0, \; s=s+i 0$, etc. are implied.

To resolve the singularity structure of Feynman integrals in $\ep$
it is very useful to apply the MB representation
\be
\frac{1}{(X+Y)^{\nu}} = \frac{1}{\Gm(\nu)}
\frac{1}{2\pi i}\int_{-i \infty}^{+i \infty} \dd z
\frac{Y^z}{X^{\nu+z}} \Gm(\nu+z) \Gm(-z) \;,
\label{MB}
\ee
that makes it
possible to  replace sums of terms raised to some power by their
products in some powers, at the cost of introducing extra
integrations.
By a straightforward generalization of two-loop manipulations \cite{ATT,S3})
one can introduce, in a suitable way,
MB integrations, first, after the integration over one of the loop
momenta, $r$, then after the integration over $l$,
and complete this procedure after integration
over the loop momentum $k$. As a result,
one arrives \cite{S5} at the following sevenfold MB representation of
(\ref{3box}):
\bea
T(a_1,\ldots,a_8;s,t ;\ep)
 &=&
\frac{\left(i\pi^{d/2} \right)^3 (-1)^a}{
\prod_{j=2,5,7,8,9,10}\Gm(a_j) \Gm(4-a_{589(10)}-2\ep)(-s)^{a-6+3\ep}}
\nn \\ &&  \hspace*{-40mm}\times
\frac{1}{(2\pi i)^7} \int_{-i\infty}^{+i\infty}
\dd w \prod_{j=2}^7 \dd z_j
\left(\frac{t}{s} \right)^{w}
\frac{\Gm(a_{2} + w)\Gm(-w)
\Gm(z_2 + z_4) \Gm(z_3 + z_4)}
{\Gm(a_1 + z_3 + z_4) \Gm(a_3 + z_2 + z_4)}
\nn \\ &&  \hspace*{-40mm}\times
\frac{
\Gm(2 - a_{12} - \ep + z_{2})
\Gm(2 - a_{23} - \ep + z_{3})
 \Gm(a_{7} + w - z_{4})  \Gm(-z_{5}) \Gm(-z_{6})
  }
{ \Gm(4 - a_{123} - 2 \ep + w - z_{4}) \Gm(a_{6} - z_{5})
  \Gm(a_{4} - z_{6}) }
\nn \\ &&  \hspace*{-40mm}\times
\Gm( + a_{123}-2 + \ep + z_{4})
 \Gm(w + z_{2} + z_{3} + z_{4} - z_{7})
\nn \\ &&  \hspace*{-40mm}\times
\Gm(2  - a_{59(10)} - \ep - z_{5} - z_{7})
  \Gm(2 - a_{589} - \ep - z_{6} - z_{7})
\nn \\ &&  \hspace*{-40mm}\times
  \Gm(a_{467}-2  + \ep + w - z_{4} - z_{5} - z_{6} - z_{7})
  \Gm(a_{9} + z_{7}) \Gm(a_{5} + z_{5} + z_{6} + z_{7})
\nn \\ &&  \hspace*{-40mm}\times
\Gm(4 - a_{467} - 2 \ep + z_{5} + z_{6} + z_{7})
 \Gm(a_{589(10)}-2+ \ep + z_{5} + z_{6} + z_{7})
\nn \\ &&  \hspace*{-40mm}\times
 \Gm(2 - a_{67} - \ep - w - z_{2} + z_{5} +
    z_{7}) \Gm(2 - a_{47} - \ep - w - z_{3} + z_{6} + z_{7})
\, ,
\label{7MB}
\eea
where $a=\sum_{i=1}^{10} a_i$, $a_{589(10)}=a_5+a_8+a_9+a_{10},
a_{123}=a_1+a_2+a_3$, etc., and
integration contours are chosen in the standard way.

In the case of the master triple box, we set $a_i=1$ for
$i=1,2,\ldots,10$ to obtain
\bea
T^{(0)}(s,t;\ep)\equiv T(1,\ldots,1;s,t;\ep) &&
\nn \\ &&  \hspace*{-50mm}
= \frac{\left(i\pi^{d/2} \right)^3}{
\Gm(-2\ep)(-s)^{4+3\ep}}
\frac{1}{(2\pi i)^7} \int_{-i\infty}^{+i\infty}
\dd w \prod_{j=2}^7 \dd z_j
\left(\frac{t}{s} \right)^{w}
\frac{ \Gm(1 + w)\Gm(-w) }{\Gm(1 - 2 \ep + w - z_4) }
\nn \\ &&  \hspace*{-50mm}\times
\frac{ \Gm(-\ep + z_2)
  \Gm(-\ep + z_3) \Gm(1 + w - z_4) \Gm(-z_2 - z_3 - z_4)
  \Gm(1 + \ep + z_4)}
{ \Gm(1 + z_2 + z_4)
  \Gm(1 + z_3 + z_4)}
\nn \\ &&  \hspace*{-50mm}\times
\frac{  \Gm(z_2 + z_4) \Gm(z_3 + z_4) \Gm(-z_5)
  \Gm(-z_6) \Gm(w + z_2 + z_3 + z_4 - z_7) }
{ \Gm(1 - z_5) \Gm(1 - z_6) \Gm(1 - 2 \ep + z_5 + z_6 + z_7) }
\nn \\ &&  \hspace*{-50mm}\times
 \Gm(-1 - \ep - z_5 - z_7) \Gm(-1 - \ep - z_6 - z_7)\Gm(1 + z_7)
\nn \\ &&  \hspace*{-50mm}\times
  \Gm(1 + \ep + w - z_4 - z_5 - z_6 - z_7)
  \Gm(-\ep - w - z_2 + z_5 + z_7)
\nn \\ &&  \hspace*{-50mm}\times
 \Gm(-\ep - w - z_3 + z_6 + z_7)
  \Gm(1 + z_5 + z_6 + z_7) \Gm(2 + \ep + z_5 + z_6 + z_7)
\, .
\label{7MB0}
\eea
Observe that, because of the presence of the factor $\Gm(-2\ep)$
in the denominator, we are forced to take some residue
in order to arrive at a non-zero result at $\ep=0$,
so that the integral is effectively sixfold.

Then the standard procedure of taking residues and shifting
contours is applied, with the goal to obtain a sum of integrals
where one may expand integrands in Laurent series in $\ep$.
One- and two-loop examples of such procedures can be found, e.g., in
\cite{S4}. The poles in $\ep$
are not visible at once, at a first integration over one of the MB
variables. However, the rule for finding a mechanism of the generation
of poles is based on the simple observation that a product of two gamma
functions
$\Gm(a+z)\Gm(b-z)$, where $z$ is a MB integration variable and $a$ and $b$
depend on the rest of the
variables, generates
a pole of the type $\Gm(a+b)$. This means that any contour in
the next integrations should be chosen according to this dependence.
So, the first step is an analysis of various pairs of gamma functions
and various orders of integration in (\ref{7MB0}).
The analysis of the integrand shows that the following four gamma functions
play a crucial role for the generation of poles in $\ep$:
$\Gm(-\ep + z_{2,3})$ and $\Gm(-1 - \ep - z_{6,5} - z_7)$.
The first decomposition of the integral (\ref{7MB0}) arises when
one either takes a residue at the first pole of one of these gamma
functions or shifts the corresponding contour, i.e. changes the
nature of this pole. As a result (\ref{7MB0}) is decomposed as
$2T_{0001}+2T_{0010}+2T_{0011}+T_{0101}+2T_{0110}+2T_{0111}+T_{1010}
+2T_{1011}+T_{1111}$
where a symmetry of the integrand is taken into account.
Here the value $1$ of an index means that a residue is taken and $0$ means a
shifting of a contour. The first two indices correspond to the
gamma functions
$\Gm(-\ep + z_{2})$ and $\Gm(-1 - \ep - z_{5} - z_7)$ and the
second two indices to $\Gm(-\ep + z_{3})$ and $\Gm(-1 - \ep - z_{6} - z_7)$,
respectively.
The term $T_{0000}$ is absent because it is zero at $\ep=0$
due to $\Gm(-2\ep)$ in the denominator.

Each of these terms is further appropriately decomposed and,
eventually, one is left with integrals where integrands can be
expanded in $\ep$. These resulting terms involve up to five
integrations. Taking some of these integrations with the help of
the first and the
second Barnes lemmas, one reduces all the integrals to no more than
twofold MB integrals of gamma functions and their
derivatives. In some of them, one more integration can be also
performed in gamma functions. Then the last integration, over $w$
is performed by taking residues and summing up
resulting series, in terms of HPL.
Keeping in mind the Regge limit, $t/s \to 0$, let us, for
definiteness, decide to close the contour of the final
integration, over $w$, to the right and obtain power series in
$t/s$.  The coefficients of these series are (up to $(-1)^n$) linear combinations
of $1/n^6, S_1(n)/n^5,\ldots, S_1 (n) S_3 (n)/n^2,\ldots$, where
$S_k(n) =  \sum_{j=1}^n j^{-k}$. Summing up these series gives
results in terms of HPL of the variable $-t/s$
which can be analytically continued to any
domain from the region $|t/s|<1$.

In the twofold MB integrals where one more integration
(over a variable different from $w$) can be
hardly performed in gamma functions, one performs it with $w$ in a vicinity
of an integer point $w=n=0,1,2,\ldots$, in expansion in $z=w-n$, with a
sufficient
accuracy. Then one obtains powers series where, in
addition to
$1/n^6, S_1(n)/n^5,\ldots$, quantities like
$S_{ik}(n) =  \sum_{j=1}^n j^{-i} S_{k}(j), S_{ikl}(n) = \sum_{j=1}^n j^{-i}
S_{kl}(j)$
appear. These series are also summed up in terms of HPL.

Eventually we arrive at the following result:
\be
T^{(0)}(s,t;\ep)=
-\frac{\left(i \pi^{d/2} \e^{-\gm_{\rm E}\ep} \right)^3 }{s^3 (-t)^{1+3\ep}}
\;
 \sum_{i=0}^6 \frac{ c_j(x,L)}{\ep^j}\;,
\ee
where $\gm_{\rm E}$ is the Euler constant, $x=-t/s$, $L=\ln(s/t)$, and
\bea
c_6=\frac{16}{9}\, , \;\;  c_5=-\frac{5}{3} L\, , \;\;
c_4=-\frac{3}{2}\pi^2 \, , &&
\nn \\ &&  \hspace*{-70mm}
c_3=
3 (H_{0, 0, 1}(x) + L H_{0, 1}(x)) + \frac{3}{2} (L^2 + \pi^2) H_{1}(x) -
\frac{11}{12} \pi^2  L - \frac{131}{9} \zeta_3\,,
\nn \\ &&  \hspace*{-70mm}
c_2=
-3 \left(17 H_{0, 0, 0, 1}(x) + H_{0, 0, 1, 1}(x) + H_{0, 1, 0, 1}(x) +
          H_{1, 0, 0, 1}(x)\right)
\nn \\ &&  \hspace*{-65mm}
- L \left(37 H_{0, 0, 1}(x)
          + 3  H_{0, 1, 1}(x) +  3  H_{1, 0, 1}(x)\right)
    - \frac{3}{2} (L^2 + \pi^2) H_{1, 1}(x)
\nn \\ &&  \hspace*{-65mm}
    - \left( \frac{23}{2} L^2 + 8 \pi^2 \right) H_{0, 1}(x)
    - \left(\frac{3}{2} L^3 + \pi^2 L - 3 \zeta_3 \right)H_{1}(x)
    + \frac{49}{3} \zeta_3 L - \frac{1411}{1080} \pi^4 \,,
\nn \\ &&  \hspace*{-70mm}
c_1=
3 \left( 81 H_{0, 0, 0, 0, 1}(x) + 41 H_{0, 0, 0, 1, 1}(x)
+ 37 H_{0, 0, 1, 0, 1}(x)
+ H_{0, 0, 1, 1, 1}(x) \right.
\nn \\ &&  \hspace*{-65mm}
+ 33 H_{0, 1, 0, 0, 1}(x) + H_{0, 1, 0, 1, 1}(x) + H_{0, 1, 1, 0, 1}(x)
+ 29 H_{1, 0, 0, 0, 1}(x)
\nn \\ &&  \hspace*{-65mm} \left.
+ H_{1, 0, 0, 1, 1}(x) + H_{1, 0, 1, 0, 1}(x) + H_{1, 1, 0, 0, 1}(x)\right)
+ L \left(177 H_{0, 0, 0, 1}(x) + 85 H_{0, 0, 1, 1}(x) \right.
\nn \\ &&  \hspace*{-65mm} \left.
+ 73 H_{0, 1, 0, 1}(x)
+ 3 H_{0, 1, 1, 1}(x) + 61 H_{1, 0, 0, 1}(x) + 3 H_{1, 0, 1, 1}(x)
+ 3 H_{1, 1, 0, 1}(x)\right)
\nn \\ &&  \hspace*{-65mm}
+ \left(\frac{119}{2} L^2 + \frac{139}{12} \pi^2\right) H_{0, 0, 1}(x)
+ \left(\frac{47}{2} L^2 + 20 \pi^2\right) H_{0, 1, 1}(x)
\nn \\ &&  \hspace*{-65mm}
+ \left(\frac{35}{2} L^2 + 14 \pi^2\right) H_{1, 0, 1}(x)
+ \frac{3}{2}\left(L^2 + \pi^2\right) H_{1, 1, 1}(x)
\nn \\ &&  \hspace*{-65mm}
+ \left(\frac{23}{2} L^3 + \frac{83}{12}\pi^2 L
- 96 \zeta_3 \right)H_{0, 1}(x)
+ \left(\frac{3}{2} L^3 + \pi^2 L - 3 \zeta_3 \right) H_{1, 1}(x)
\nn \\ &&  \hspace*{-65mm}
+ \left(\frac{9}{8} L^4 + \frac{25}{8} \pi^2 L^2 - 58 \zeta_3  L
+ \frac{13}{8} \pi^4\right) H_{1}(x)
- \frac{503}{1440}\pi^4 L
+ \frac{73}{4} \pi^2 \zeta_3  - \frac{301}{15} \zeta_5\,,
\nn \\ &&  \hspace*{-70mm}
c_0=
-\left( 951 H_{0, 0, 0, 0, 0, 1}(x) + 819 H_{0, 0, 0, 0, 1, 1}(x)
+ 699 H_{0, 0, 0, 1, 0, 1}(x) + 195 H_{0, 0, 0, 1, 1, 1}(x)
\right.
\nn \\ &&  \hspace*{-65mm}
+ 547 H_{0, 0, 1, 0, 0, 1}(x) + 231 H_{0, 0, 1, 0, 1, 1}(x)
+ 159 H_{0, 0, 1, 1, 0, 1}(x) + 3 H_{0, 0, 1, 1, 1, 1}(x)
\nn \\ &&  \hspace*{-65mm}
+ 363 H_{0, 1, 0, 0, 0, 1}(x) + 267 H_{0, 1, 0, 0, 1, 1}(x)
+ 195 H_{0, 1, 0, 1, 0, 1}(x) + 3 H_{0, 1, 0, 1, 1, 1}(x)
\nn \\ &&  \hspace*{-65mm}
+ 123 H_{0, 1, 1, 0, 0, 1}(x) + 3 H_{0, 1, 1, 0, 1, 1}(x)
+ 3 H_{0, 1, 1, 1, 0, 1}(x) + 147 H_{1, 0, 0, 0, 0, 1}(x)
\nn \\ &&  \hspace*{-65mm}
+ 303 H_{1, 0, 0, 0, 1, 1}(x) + 231 H_{1, 0, 0, 1, 0, 1}(x)
+ 3 H_{1, 0, 0, 1, 1, 1}(x) + 159 H_{1, 0, 1, 0, 0, 1}(x)
\nn \\ &&  \hspace*{-65mm}
+ 3 H_{1, 0, 1, 0, 1, 1}(x) + 3 H_{1, 0, 1, 1, 0, 1}(x)
+ 87 H_{1, 1, 0, 0, 0, 1}(x) + 3 H_{1, 1, 0, 0, 1, 1}(x)
\nn \\ &&  \hspace*{-65mm} \left.
+ 3 H_{1, 1, 0, 1, 0, 1}(x) + 3 H_{1, 1, 1, 0, 0, 1}(x)
\right)
\nn \\ &&  \hspace*{-65mm}
- L \left(729 H_{0, 0, 0, 0, 1}(x) + 537 H_{0, 0, 0, 1, 1}(x)
  + 445 H_{0, 0, 1, 0, 1}(x) + 133 H_{0, 0, 1, 1, 1}(x)\right.
\nn \\ &&  \hspace*{-65mm}
  + 321 H_{0, 1, 0, 0, 1}(x) + 169 H_{0, 1, 0, 1, 1}(x)
  + 97 H_{0, 1, 1, 0, 1}(x) + 3 H_{0, 1, 1, 1, 1}(x)
\nn \\ &&  \hspace*{-65mm}
  + 165 H_{1, 0, 0, 0, 1}(x) + 205 H_{1, 0, 0, 1, 1}(x)
  + 133 H_{1, 0, 1, 0, 1}(x) + 3 H_{1, 0, 1, 1, 1}(x)
\nn \\ &&  \hspace*{-65mm} \left.
  + 61 H_{1, 1, 0, 0, 1}(x) + 3 H_{1, 1, 0, 1, 1}(x)
  + 3 H_{1, 1, 1, 0, 1}(x)\right)
\nn \\ &&  \hspace*{-65mm}
- \left(\frac{531}{2} L^2 + \frac{89}{4} \pi^2\right) H_{0, 0, 0, 1}(x)
- \left(\frac{311}{2} L^2 + \frac{619}{12}\pi^2\right) H_{0, 0, 1, 1}(x)
\nn \\ &&  \hspace*{-65mm}
- \left(\frac{247}{2} L^2 + \frac{307}{12} \pi^2\right) H_{0, 1, 0, 1}(x)
- \left(\frac{71}{2} L^2 + 32 \pi^2\right) H_{0, 1, 1, 1}(x)
\nn \\ &&  \hspace*{-65mm}
- \left(\frac{151}{2} L^2 - \frac{197}{12} \pi^2\right) H_{1, 0, 0, 1}(x)
- \left(\frac{107}{2} L^2 + 50 \pi^2\right) H_{1, 0, 1, 1}(x)
\nn \\ &&  \hspace*{-65mm}
- \left(\frac{35}{2} L^2 + 14 \pi^2\right) H_{1, 1, 0, 1}(x)
- \frac{3}{2}\left( L^2 +   \pi^2\right) H_{1, 1, 1, 1}(x)
\nn \\ &&  \hspace*{-65mm}
- \left(\frac{119}{2} L^3 + \frac{317}{12} \pi^2 L
- 455 \zeta_3 \right)H_{0, 0, 1}(x)
- \left(\frac{47}{2} L^3 + \frac{179}{12} \pi^2  L
- 120 \zeta_3 \right)H_{0, 1, 1}(x)
\nn \\ &&  \hspace*{-65mm}
- \left(\frac{35}{2} L^3 + \frac{35}{12} \pi^2  L
- 156 \zeta_3 \right)H_{1, 0, 1}(x)
- \left(\frac{3}{2} L^3 + \pi^2 L - 3 \zeta_3 \right)H_{1, 1, 1}(x)
\nn \\ &&  \hspace*{-65mm}
- \left(\frac{69}{8} L^4 + \frac{101}{8} \pi^2 L^2
- 291 \zeta_3 L + \frac{559}{90} \pi^4 \right)H_{0, 1}(x)
\nn \\ &&  \hspace*{-65mm}
- \left(\frac{9}{8} L^4 + \frac{25}{8} \pi^2 L^2 - 58 \zeta_3 L
+ \frac{13}{8} \pi^4\right) H_{1, 1}(x)
\nn \\ &&  \hspace*{-65mm}
- \left(\frac{27}{40} L^5 + \frac{25}{8} \pi^2 L^3
- \frac{183}{2} \zeta_3  L^2
   + \frac{131}{60} \pi^4 L - \frac{37}{12} \pi^2 \zeta_3
   + 57 \zeta_5 \right) H_{1}(x)
\nn \\ &&  \hspace*{-65mm}
+ \left(\frac{223}{12}  \pi^2 \zeta_3  + 149 \zeta_5 \right) L
+ \frac{167}{9} \zeta_3^2 - \frac{624607}{544320} \pi^6 \,.
\label{3box-result}
\eea
Here $\zeta_3=\zeta(3),\zeta_5=\zeta(5)$ and
$\zeta(z)$ is the Riemann zeta function.
The functions $H_{a_1,a_2,\ldots,a_n}(x)\equiv H(a_1,a_2,\ldots,a_n;x)$,
with $a_i=1,0,-1$, are
HPL \cite{HPL} which are recursively defined by
\[
H(a_1,a_2,\ldots,a_n;x) = \int_0^x f(a_1;t) H(a_2,\ldots,a_n;t)\,,
\]
where
\[f(\pm 1;x)=1/(1 \mp x),\;\;f(0;x)=1/x,\;\;
H(\pm 1;x)= \mp \ln(1\mp x),\;\; H(0;x)= \ln x \,.\]

In (\ref{3box-result}), only HPL with parameters $0$ and $1$
are involved. If a given HPL involves only parameters $a_i=0$ and $1$
and the number of these parameters is less or equal to
four, it can be expressed \cite{HPL} in terms of
usual polylogarithms Li$(x)$ \cite{Lewin} and generalized polylogarithms
\cite{GenPolyLog}
\[
  S_{a,b}(x) = \frac{(-1)^{a+b-1}}{(a-1)! b!}
    \int_0^1 \frac{\ln^{a-1}(t)\ln^b(1-x t)}{t} \dd t \,.
\]
(See \cite{S6} where the coefficients $c_j$, with $j\geq2$ are
expressed in terms of (generalized) polylogarithms.)

The above result was confirmed \cite{BH1}
with the help of  numerical integration in the space of
alpha parameters \cite{BH}.
Another natural check of the result is its agreement with the leading
power Regge asymptotic behaviour \cite{S5} which was evaluated by
an independent method based on the strategy of expansion by
regions \cite{BS}.

The procedure described above can be applied, in a similar way, to the
calculation of any massless planar on-shell triple box.
At a first step, one has to take care of the following four gamma functions in
(\ref{7MB}):
\[
\Gm(2 - a_{12} - \ep + z_{2}),
\Gm(2 - a_{23} - \ep + z_{3}), \Gm(2  - a_{59(10)} - \ep - z_{5} -
z_{7}),
  \Gm(2 - a_{589} - \ep - z_{6} - z_{7}) \;.
\]
This procedure gives a decomposition similar to $2T_{0001}+2T_{0010}+\ldots$.
Next steps will be also generalizations of the corresponding steps
in the evaluation of (\ref{7MB0}).
Hopefully, such a procedure can be made automatic by means of
computer algebra.

The result presented above shows that analytical
calculations of four-point on-shell massless Feynman diagrams
at the three-loop
level are quite possible so that one may think of evaluating
three-loop virtual corrections to various scattering processes.

\vspace{0.5 cm}

{\em Acknowledgments.}
I am very much grateful to G.~Heinrich for numerical checks.
This work was supported by the Russian Foundation for Basic
Research through project 01-02-16171, INTAS through grant 00-00313,
and the Volkswagen Foundation through contract No.~I/77788.

\end{document}